\documentclass{aastex631}

\usepackage{breqn}
\usepackage{amsmath}

\begin{document}

\title{Neutrino Oscillation Effects on the Luminosity of Neutrino-Dominated Accretion Flows Around Black Holes}

\correspondingauthor{Poemwai Chainakun}
\email{pchainakun@g.sut.ac.th}

\author{Chitipat Deesamer}
\affiliation{School of Physics, Institute of Science, Suranaree University of Technology, Nakhon Ratchasima 30000, Thailand}

\author[0000-0002-9099-4613]{Poemwai Chainakun}
\affiliation{School of Physics, Institute of Science, Suranaree University of Technology, Nakhon Ratchasima 30000, Thailand}
\affiliation{Center of Excellence in High Energy Physics and Astrophysics, Suranaree University of Technology, Nakhon Ratchasima 30000, Thailand}

\author{Warintorn Sreethawong}
\affiliation{School of Physics, Institute of Science, Suranaree University of Technology, Nakhon Ratchasima 30000, Thailand}
\affiliation{Center of Excellence in High Energy Physics and Astrophysics, Suranaree University of Technology, Nakhon Ratchasima 30000, Thailand}

\begin{abstract}

The stellar-mass black hole surrounded by the neutrino-dominated accretion flow (NDAF) is proposed to be the central engine of the gamma-ray burst (GRB). In this work, we investigate the neutrino/anti-neutrino luminosity and annihilation luminosity from the NDAF and pair-annihilation model, taking into account the neutrino oscillation above the accretion disk. The disk hydrodynamical properties are modelled using the empirical solution previously derived with boundary conditions including the effect of electron degeneracy and neutrino trapping. Our key parameters are the mass accretion rate and the black hole spin given in the range of $\dot{M} = 0.1$--10~$M_\odot$~s$^{-1}$ and $ 0 \le a < 1$, respectively. Without neutrino oscillation, the obtained neutrino/anti-neutrino luminosity is $\sim 10^{51}$--$10^{53}~\textrm{erg s}^{-1}$, while the neutrino annihilation luminosity is found to be $\sim 10^{46}$--$10^{51}~\textrm{erg s}^{-1}$. In the presence of neutrino oscillation in the vacuum limit, the electron neutrino annihilation luminosity decreases by $\lesssim 22\%$ through the flavor transformation, while the muon- and tau-neutrino luminosity can increase up to $\sim 45\%$ and $ 60\%$, respectively. As a result, the total annihilation luminosity can be reduced up to $\sim 19\%$ due to the oscillation process above the disk. Finally, we also investigate the case whereby the CP-violating phase is changed from $\delta \textrm{CP} =$ $0^{\circ}$ to $245^{\circ}$. However, our results reveal that the CP-violating phase has minimal impact on neutrino annihilation luminosity.

\end{abstract}

\keywords{accretion disks ---  black hole physics --- neutrinos}

\section{Introduction} \label{sec:intro}
Neutrinos, being weakly interacting subatomic particles, present formidable challenges in their detection. Previous studies have shown that neutrinos can be released from a variety of sources including nuclear reactions in laboratories \citep{ABE2011106,AN201278,Li:2014qca}, atmospheric processes \citep{Atmos}, and astrophysical phenomena including solar emissions, supernovae \citep{IceCube,GERHARDT201085,BELLERIVE201630,Yokoyama:2017mnt} and gamma-ray burst (GRB) \citep{GRBPac,KohriNDAF2002,GRBIceCube}. Among these sources, GRBs
are notable for being the most energetic explosions, emitting a tremendous amount of energy on the  order of $10^{51}$--$10^{53}$ erg~s$^{-1}$ within a wide range of timescale. Short GRBs last merely a second, while long GRBs can persist for several hundred seconds \citep{GRB,GRBhistory,GRBLUM1,SGRBLGRB}.

A neutrino-dominated accretion flow (NDAF) around rotating black holes has been proposed as a plausible candidate for the central engine of GRBs \citep{PophamAccre,NaraAccre,KohriNDAF2005,ChenNDAF,XueNDAF,LiuNDAF,NagatakiNDAF,ChenVerti,WeiNDAF2}. The disk could originate from various scenarios, such as the merger of double neutron star binaries (NS-NS) \citep{NSNS}, neutron star-black hole binaries (NS-BH) \citep{NSBH}, and the collapse of massive stars \citep{massivestar}. The NDAF has an extremely high temperature of $T \sim 10^{10}$--$10^{11}~\textrm{K}$ and a high density of $\rho \sim 10^{7}$--$10^{13}$~g~cm$^{-3}$ at the inner-disk region \citep{PophamAccre}, where a large number of neutrinos and anti-neutrinos with energy of $\sim 10$--30~MeV is produced through the cooling processes in the flows and emit from the surface of the accretion disk \citep{LiuDetect2016}. In the context of BH mergers (with an evolutionary timescale of $\sim$ 1--100 s), neutrino emission could yield a neutrino luminosity up to $\sim$ 10$^{53}$--10$^{55}$ erg s$^{-1}$ \citep{PophamAccre,XueNDAF}. Above the disk, some fractions of neutrino and anti-neutrino could annihilate and produce annihilation luminosity approximately one to two orders of magnitude smaller. Nevertheless, this luminosity is sufficient to generate relativistic electron-positron pairs to drive a strong outflow like a fireball, hence enabling the formation of a GRB \citep{AnniGRB}. Meanwhile, there were investigations of the GRB-driven mechanism considering magnetohydrodynamics (e.g., the Blandford-Znajek mechanism), which also contribute to neutrino emission and neutrino annihilation \citep{BlandMagne,McMagne,HawleyMagne,KawaMagne,NoriMagne,SheMagne}. 

Up until now,  the NDAF model has been extensively studied. The pioneering work by \cite{PophamAccre} is the first investigation of the neutrino-cooled accretion disk around a Kerr black hole taking into account general relativistic effects with the thin-disk approximation for common mass accretion rates of stellar-mass black holes ($\dot{M} \sim$ 0.01--10 $M_{\odot}$ s$^{-1}$). Notably, the neutrino trapping effect arising as a consequence of opacity could change the disk regions where neutrinos are emitted \citep{DiMatNDAF,GuNDAF}. The electron degeneracy parameter was introduced into the model, influencing both the disk structure and neutrino emission \citep{KohriNDAF2002,KohriNDAF2005}. Later, \cite{XueNDAF} presented the global solution for NDAF, considering the effect of electron degeneracy and mass fraction (see \citealt[][for a review]{LiuNDAF,NagatakiNDAF}). The possibility of detecting MeV neutrinos from NDAF models using multi-messenger approaches has also been explored \citep{PanVerti,LiuVerti2,CabalNDAF,LiuDetect2016,WeiNDAF}. Based on the Hyper-Kamiokande detector (Hyper-K), neutrino detection was approximated by evaluating the neutrino and anti-neutrino luminosities from GRB-related events in various galaxies at distances of $\sim$ 0.61--0.77 Mpc, with a total observed energy of $\sim$ 10$^{52}$ erg and an estimated detection rate of $\sim$ 0.10--0.25 events per century \citep{LiuDetect2016}. 

However, it is essential to consider that neutrinos can undergo oscillations and transform into other flavors as they propagate \citep{NeuOscFirst,NeuAtmos,NeuSolar}. This phenomenon has the potential to significantly influence the luminosity profiles of NDAF by altering the neutrino flavor distribution over time. The effect of neutrino oscillation has been investigated in the context of merger events, where oscillations may enhance r-process nucleosynthesis in the outflow \citep{OscGen,OscNN1,OscNN2,OscNN3,OscNN4}. Despite these efforts, the full impact of neutrino oscillations on neutrino luminosity remains incompletely understood. By incorporating neutrino oscillation effects into the NDAF model, \cite{UribeOsc} investigated the neutrino oscillation inside the accretion disk and found that the total luminosity and annihilation luminosity can be decreased by a factor of 4--5. They also discovered that the oscillation above the disk, which is still neglected in their work, should be considered at the same time.

In this work, we investigate the impact of neutrino oscillation on the annihilation luminosity above the disk within the NDAF framework. We adopt the thin-disk approximation \citep{PophamAccre} and simplify the NDAF model following previous assumptions \citep{KohriNDAF2005,ChenNDAF,LiuDetect2016,LiuNDAF}. In Section \ref{sec:model}, the disk equations and assumptions of the NDAF model are outlined. In Section \ref{sec:oscillation}, the approach of neutrino oscillation above the disk as well as the annihilation luminosity calculation is described. The predicted disk profiles and corresponding annihilation luminosity are presented in Section \ref{sec:result}. Furthermore, we also consider the neutrino oscillation in two cases including CP-violating phase $\delta \textrm{CP} = 0^{\circ}$ and $245^{\circ}$, which is a recent plausible value from experiments \citep{PDG}. Finally, we discuss the implications of neutrino oscillation above the disk and its potential impact on the observed annihilation luminosity in Section \ref{sec:con&discussion}.

\section{Neutrino-Dominated Accretion Flow Model} \label{sec:model}

\subsection{Geometry and Model Assumptions} \label{subsec:modelreg}

\begin{figure*}[ht]
\centerline{
\includegraphics[width=0.8\textwidth]{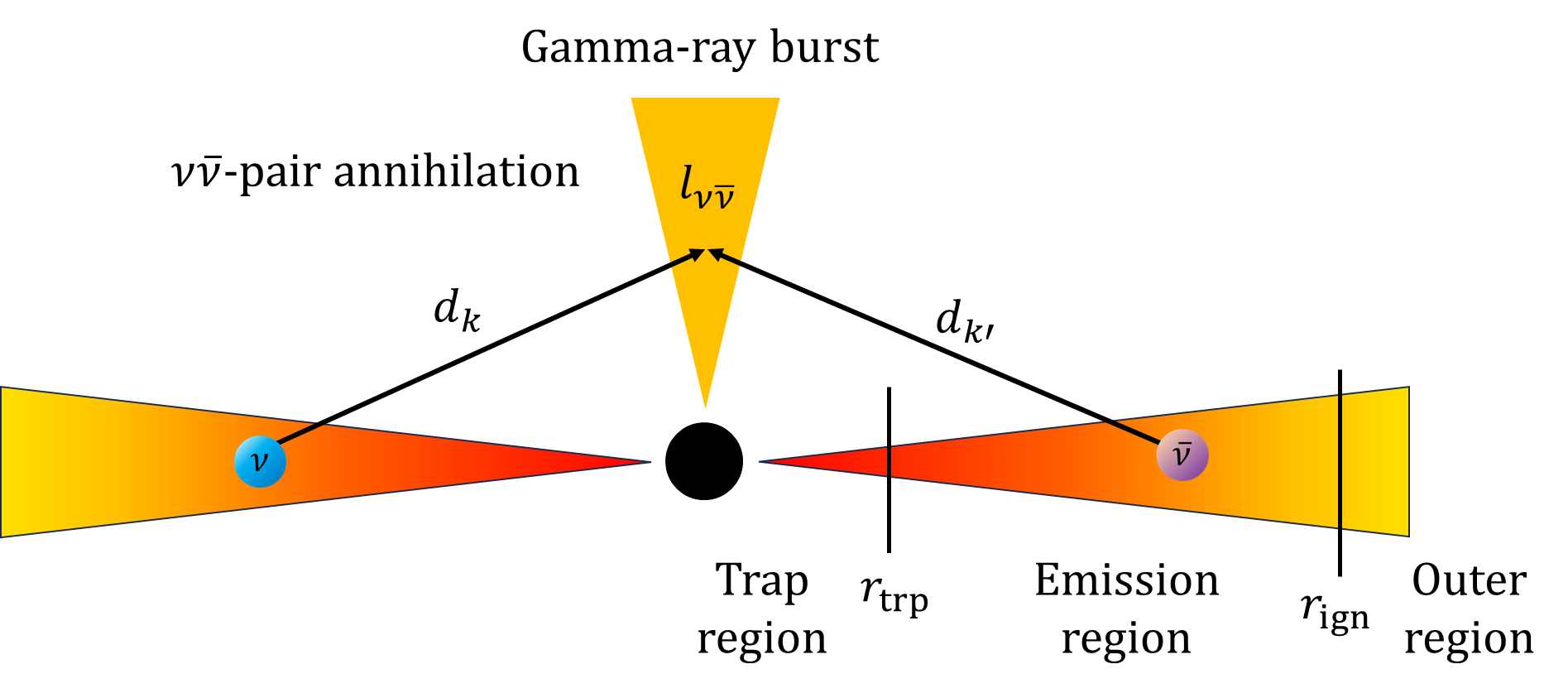}
}
\caption{Schematic of a simplified neutrino-cooled accretion disk model (not to scale). The disk regions are separated into the trapping region ($r < r_{\textrm{trp}})$, emission region ($r_{\textrm{trp}}< r < r_{\textrm{ign}}$), and outer region ($r > r_{\textrm{ign}}$). Neutrino $\nu$ and anti-neutrino $\bar{\nu}$ emitted from the surface of the disk will propagate through space with distances $d_{k}$ and $d_{k'}$ respectively, and will annihilate above the disk yielding annihilation luminosity $l_{\nu\bar{\nu}}$ at specific points.}
\label{fig:1}
\end{figure*}

Firstly, we employ the NDAF model presented in \cite{PophamAccre}. The accretion disk is assumed to be in steady-state where the accretion timescale is significantly shorter than the evolution timescale of mass accretion rate, and hence the disk can be characterized into distinctly separated regions \citep{ChenNDAF}, as illustrated in Figure~\ref{fig:1}. We adopt the thin-disk approximation, where $H/r \le$ 0.1--0.4 ($H$ is the half-thickness of the disk at radius $r$ measured from the central black hole). The key parameters governing the hydrodynamic characteristics of the flow consist of the black hole spin ($a$) and the mass accretion rate ($\dot{M}$). We set the outer radius of the disk at $r_{\textrm{out}}=500$~$r_{\rm g}$ and the black hole mass $M =$ 3~$M_{\odot}$ (1~$r_{\rm g} = 2~GM/c^{2}$ , where $G = 6.67\times10^{-8} \textrm{ erg cm$^{-3}$ s$^{-1}$}$ is gravitational constant and $c = 3\times10^{10} \textrm{ cm s$^{-1}$}$ is the speed of light) to ensure coverage of the emission radius \citep{XueNDAF}. For the innermost region near the black hole, the advection effect dominates and the flows become optically thick. Consequently, photons and neutrinos/anti-neutrinos are trapped, precluding radiation emission. The trapping radius is chosen to be $r_{\textrm{trp}}$= 1.473 $r_{\textrm{g}}$ which is an intermediate value and found in the cases of $ 0 \le a < 1$ and $\dot{M} \sim$ 0.01--10 $M_{\odot}$ s$^{-1}$ \citep{XueNDAF}. For comparison, we also look at the case when $r_{\textrm{trp}}$ = 3 $r_{\textrm{g}}$ in order to see the effect of variable $r_{\textrm{trp}}$. Within the ignition, or emission region ($r_{\rm trp} < r < r_{\textrm{ign}}$), the balancing between the viscous heating and the cooling rate is satisfied so that neutrino emission switches on \citep{ChenNDAF,ZalaIGN}. 

Under this setup, the neutrino and anti-neutrino emissions from the cooling process within the ignition region, $r_{\textrm{trp}} < r < r_{\textrm{ign}}$, can be calculated. Typically, the emission region extends largely to cover the area where neutrons and protons dominate the disk before transitioning into a region dominated by heavy atoms through the nucleosynthesis in the outermost region \citep{MeyerNSE,XueNDAF,JaniukNSE,SheMagne}, so we assume $r_{\rm ign} = r_{\textrm{out}}$. Subsequently, the neutrinos and anti-neutrinos emitted from the surface of the disk propagate through space covering distances $d_{k}$ and $d_{k'}$ respectively (as depicted by arrows in Figure~\ref{fig:1}) before they annihilate. The annihilation luminosity above the disk, denoted as $l_{\nu\bar{\nu}}$, are computed for each point. Then, we compute the obtained total luminosity, comparing between the cases with and without the effects of the neutrino oscillation, as explained in the following sections.

\subsection{Disk Temperature and Density Profile}

The temperature $T$ and the density $\rho$ of the accretion disk can be determined by numerically solving the equation of state of the disk that involves considering continuity equations and balancing the cooling mechanism with various thermodynamic properties such as internal energy and pressure \citep{PophamAccre,KohriNDAF2005,ChenNDAF,XueNDAF,LiuNDAF}. Additionally, the temperature and density profiles depend on the choice of model assumptions. The inclusion of chemical processes within the flow such as neutrino trapping effects \citep{DiMatNDAF}, electron degeneracy in neutrino emission \citep{KohriNDAF2005}, and photodisintegration for heavy nuclei from nuclear statistical equilibrium (NSE) \citep{SeitenNSE}, all contribute to the cooling mechanism and result in changes to the thermodynamic properties of the disk. The relativistic calculation of the hydrodynamic properties of the NDAF, incorporating detailed neutrino physics, the balance of chemical potentials, photodisintegration, and nuclear statistical equilibrium, was conducted by \cite{XueNDAF}. Later, \cite{LiuDetect2016} derived solutions for the disk temperature, which depend on key parameters including $a$ and $\dot{M}$.

In this work, we employ the empirical solution from \cite{LiuDetect2016} to introduce the temperature profile of the disk: 
\begin{equation}
    \log{T} ({\rm K}) = 11.09 + 0.10a + 0.20\log{\dot{m}} - 0.59\log{R}  \;,
\label{eq:1}
\end{equation}
where $\dot{m} = \dot{M}/ (M_{\odot}$ s$^{-1})$ and $R = r/r_{\rm g}$. 
%
%
To determine the density profile of the disk, we consider the charge-neutrality constraints for the neutral disk \citep{LiuNDAF2007}. Specifically, the number density of protons $n_p$ must equal to the number density of electrons $n_{e^{-}}$ excluding the positron pairs $n_{e^{+}}$, i.e. 
\begin{equation}
    n_p  = \frac{\rho Y_{e}}{m_{p}} = n_{e^{-}}- n_{e^{+}}, 
\label{eq:3}
\end{equation}
where $\rho$ and $m_{p}$ are the density and mass of the proton. $Y_{e}$ is the proton-to-nucleon ratio (related to proton-neutron number density). In this case, we assume that the matter is completely dissociated where nuclear statistical equilibrium (NSE) is neglected \citep{MeyerNSE} and thus the atomic mass fraction can also be neglected \citep{KohriNDAF2005}. Then, one can obtain $Y_e$ in terms of the proton number density $n_{p}$ and the neutron number density $n_{n}$ as
\begin{equation}
    Y_{e} = \frac{n_{p}}{n_{p}+n_{n}}.
\label{eq:2}
\end{equation}
In the equilibrium radiation reaction $e^{+} + e^{-} \leftrightarrow \gamma + \gamma$, the chemical potential for photons is zero. It follows that $\mu_{e^{-}} + \mu_{e^{+}} = 0$ and we denote the electron chemical potential $\mu_{e^{-}} = \mu_{e}$. The number density of electron/positron is given by the Fermi-Dirac integral,
\begin{equation}
    n_{e^{\pm}} = \frac{1}{\hbar^3\pi^2} \int^{\infty}_{0}{\frac{1}{e^{(\sqrt{p^2c^2+m^2_{e}c^4}\mp\mu_{e})/k_{B}T}+1} p^2 dp}.
\label{eq:4}
\end{equation}
In perfect equilibrium, the precise calculation of the electron chemical potential is determined by the reaction between protons and neutrons \citep{KohriNDAF2005},
\begin{equation}
    \left( \frac{n}{p}\right)_{eq} = \exp\left( -\frac{Q}{k_{B}T}+\eta_{e}\right),
\label{eq:5}
\end{equation}
where $\eta_{e} = \mu_{e}/k_{B}T$ is the degeneracy of electrons and $Q  = $ $(m_{n}-m_{p})c^{2} \simeq 1.29 \rm~MeV$  is the rest-mass difference between a neutrino and a proton. For the conditions in Equations (\ref{eq:3}) and (\ref{eq:5}), we simplify the thermodynamic quantities by fixing the number of neutrino degeneracy parameter $\eta_{e}$ for all disk emission radii with the boundary condition $Y_{e} = 0.5$ at the outer radius \citep{LiuNDAF}. The number density of each particle: $n_{e^-}$, $n_{e^+}$, $n_p$, $n_n$, the electron fraction $Y_{e}$, and the matter density $\rho$ can be estimated in the simplest way instead of solving  complicated non-linear equations. The obtained parameters $T$ and $\rho$ are then used to determine the cooling rate and the luminosity of the disk.

\subsection{Cooling Processes and Luminosity} \label{subsec:text}
Different from other accretion disk models, NDAF has the cooling mechanism that provides neutrino emission dominating the disk. The cooling rate due to the neutrino loss in the cooling process is given by \citep{DiMatNDAF,KohriNDAF2005,LiuNDAF2007},
\begin{equation}
    Q_{\nu} = \sum_{i}Q_{\nu_{i}} = \sum_{i}{\frac{(7/8)\sigma T^{4}}{(3/4)[\tau_{\nu_{i}}/2 + 1/\sqrt{3} + 1/(3\tau_{a,\nu_{i}}) ]}},
\label{eq:6}
\end{equation}
where $\sigma$ is the Stephan-Boltzmann constant. The subscript $i$ runs for three flavors of neutrinos: electron neutrino ($\nu_{e}$), muon neutrino ($\nu_{\mu}$), and tau neutrino ($\nu_{\tau}$). The neutrino optical depth $\tau_{\nu}$ is defined by
\begin{equation}
    \tau_{\nu_{i}} = \tau_{s,\nu_{i}} + \tau_{a,\nu_{i}},
\label{eq:7}
\end{equation}
where $\tau_{s,\nu_{i}}$ and $\tau_{a,\nu_{i}}$ represent the neutrino optical depths resulting from scattering and absorption, respectively. These quantities can be expressed as
\begin{equation}
    \tau_{s,\nu_{i}} = H(\sigma_{e,\nu_{i}}n_{e} + \sum_{j}{\sigma_{j,\nu_{i}}n_{j}}),
\label{eq:8}
\end{equation}
\begin{equation}
    \tau_{a,\nu_{i}} = \frac{q_{\nu_{i}} H}{4(7/8)\sigma T^{4}}.
\label{eq:9}
\end{equation}
Here, $\sigma_{e,\nu_{i}}$ and $\sigma_{j,\nu_{i}}$ denote the cross sections of electron and nucleons, respectively. $n_{e}$ and $n_{j}$~$(j = 1, 2, 3, ...)$ are the number densities of electrons and nucleons, respectively. Note that $n_{1}$ and $n_{2}$ are the number density of free protons and free neutrons. $q_{\nu_{i}}$ is the total neutrino cooling rate (per unit volume) encompassing four main cooling processes: Urca process, electron-positron annihilation, Bremsstrahlung, and Plasmon decay. Further details regarding the cross-section calculations are provided in Appendix~\ref{sec:neutrinoemis}. 

The neutrino luminosity can be obtained by integrating the neutrino cooling rate $Q_{\nu}$ along one-dimensional radius,
\begin{equation}
    L_{\nu} = 4\pi \int^{r_{\textrm{out}}}_{r_{\textrm{in}}} Q_{\nu}r^{'}dr^{'} ,
\label{eq:10}
\end{equation}
where the inner radius for the neutrino emission is set to be $r_{\textrm{in}} = r_{\textrm{trp}} =$ 1.473 $r_{\textrm{g}}$, while the outer radius is set to be $r_{\textrm{out}} = r_{\textrm{ign}} = 500~r_{\textrm{g}}$. The value $4\pi$ refers to the radiation emitted from all directions in spherical symmetry. 

\section{Neutrino oscillation} \label{sec:oscillation}

\subsection{Annihilation Luminosity} \label{subsec:annihilation}

To calculate the neutrino annihilation luminosity from the NDAF model, we follow the non-relativistic approach outlined in \cite{PophamAccre}, \cite{RuffertNDAF}, \cite{RosswogNDAF}, \cite{LiuNDAF2007}, and \cite{KawaMagne}. We neglect the general relativistic effect (e.g., path bending and redshift) to reduce the complexity of neutrino trajectories. However, this effect should reduce the luminosity by less than a factor of 10 due to the geometry of the emitting surface near a black hole \citep{PophamAccre,Rela01,Rela02}.

In our calculation, we create a grid of cells for the disk in the equatorial plane by dividing the disk logarithmically into 400 bins. A cell $k$ ($k'$) on the disk has its mean neutrino (anti-neutrino) energy $\epsilon^{k}_{\nu_{i}}$ ($\epsilon^{k'}_{\bar{\nu}_{i}}$) and differential neutrino luminosity $l^{k}_{\nu_{i}}$ ($l^{k'}_{\bar{\nu}_{i}}$) for each neutrino (anti-neutrino) flavor $\nu_{i}$ ($\bar{\nu}_{i}$). We calculate $l^{k}_{\nu_{i}} = Q_{\nu_i} r_{k} \Delta{r_{k}}$ by integrating the cooling rate (Equation \ref{eq:6}) of each flavor of neutrino $Q_{\nu_i}$ separately over the surface of cell $k$ before summing the luminosity over the entire disk surface. Similarly, we calculate $l^{k'}_{\bar{\nu}_{i}}$ for the anti-neutrino. Note that $r_{k}$ is the distance from the central black hole to the center of the cell $k$, and $\Delta r_{k}$ is the radial size of cell $k$. Let us denote $d_k$ as the distance traveled by neutrinos emitted from a specific cell $k$ on the disk to the annihilation point, and $d_{k'}$ as the corresponding distance for anti-neutrinos emitted from another cell $k'$. The neutrino annihilation luminosity at each specific point above the disk is given by summing the contributions of neutrino and anti-neutrino radiation from all pairs of disk cells \citep{PophamAccre},

\begin{equation*}
    l_{\nu\bar{\nu}} = \sum_{i}{A_{1,i}\sum_{k}{\frac{l^{k}_{\nu_{i}}}{d^{2}_{k}}\sum_{k'}{\frac{l^{k'}_{\bar{\nu}_{i}}}{d^{2}_{k'}}}(\epsilon^{k}_{\nu_{i}} + \epsilon^{k'}_{\bar{\nu}_{i}})(1-\cos{\theta_{kk'}})^{2}}}
\end{equation*}
\begin{equation}
    + \sum_{i}{A_{2,i}\sum_{k}{\frac{l^{k}_{\nu_{i}}}{d^{2}_{k}}\sum_{k'}{\frac{l^{k'}_{\bar{\nu}_{i}}}{d^{2}_{k'}}}\frac{\epsilon^{k}_{\nu_{i}} + \epsilon^{k'}_{\bar{\nu}_{i}}}{\epsilon^{k}_{\nu_{i}}\epsilon^{k'}_{\bar{\nu}_{i}}}(1-\cos{\theta_{kk'}})}},
\label{eq:11}
\end{equation}
where $A_{1,i} = (1/12\pi^{2})[\sigma _{0}/c(m_{e}c^{2})^{2}][(C_{V,\nu_{i}} - C_{A,\nu_{i}})^2 + (C_{V,\nu_{i}} + C_{A,\nu_{i}})^2]$, $A_{2,i} = (1/6\pi^{2})[\sigma_ {0}/c][(2C^{2}_{V,\nu_{i}} - C^2_{A,\nu_{i}})]$, and $\theta_{kk'}$ is the angle at which neutrinos from cell $k$ encounter anti-neutrinos from another cell $k'$. Here, $d$ and $\epsilon$ are in cm and erg units, respectively. All constant values are shown in Appendix \ref{sec:neutrinoemis}.

The total neutrino annihilation luminosity, $L_{\nu\bar{\nu}}$, is obtained by integrating over the whole space outside the black hole and the disk \citep{XueNDAF},
\begin{equation}
    L_{\nu\bar{\nu}} = 4\pi\int^{\infty}_{H}\int^{\infty}_{r_{\textrm{in}}}l_{\nu\bar{\nu}}(r',z')r'dr'dz',
\label{eq:12}
\end{equation}
where $r_{\textrm{in}} = r_{\textrm{trp}}$.

\subsection{Neutrino Oscillation} \label{subsec:interpretation}
We estimate the annihilation luminosity from neutrinos and anti-neutrinos ejected from the disk's surface to the space above the disk. In this model, this emission causes an outflow that consists of a large number of neutrinos and anti-neutrinos, so the interaction between neutrino-neutrino before annihilation should be inevitable. Therefore, this can undergo the flavor oscillation under matter and self-interaction \citep{OscGen,MatterS}. However, this interaction may affect only the region near a black hole due to the outstanding number of neutrinos and anti-neutrinos produced in the inner area of the disk with dense matter and hot temperatures. In our calculation, we neglect the interaction between the neutrino itself in the outflow and focus on the annihilation in the whole space. This work aims to show the effect of vacuum oscillation only, which was not explicitly calculated before for this system.

We calculate the neutrino oscillation with three neutrino flavor eigenstates ($\nu_e$, $\nu_\mu$, $\nu_\tau$) and their antiparticles, initially produced in the cooling mechanism via weak interaction. We assume that they are emitted from the disk simultaneously. Subsequently, they propagate out of the disk with their mass eigenstates:
\begin{equation}
    | \nu_{i} \rangle = \sum_{\alpha=1,2,3}U_{i \alpha }| \nu_{\alpha} \rangle,
\label{eq:13}
\end{equation}
where $U_{i \alpha}$ is the Pontecorvo-Maka-Nakagawa-Sakata (PMNS) matrix,  a unitary matrix describing neutrino mixing (see Appendix \ref{sec:oscparams} for full details).  These neutrinos and anti-neutrinos oscillate until they reach the specific points above the disk and annihilate. The traveling distance of neutrino and anti-neutrino are approximated by $d_k$ and $d_{k'}$, which play a role in determining neutrino oscillation behavior as they traverse through space before annihilating. For simplicity, we further assume that the three neutrino (anti-neutrino) flavors from the same cell $k$ have neither self-interaction nor interaction with each other, and the energies carried by these particles are the same at the annihilation point.
Considering that most of the annihilation points are far enough from the black hole and assuming there is no outflow from the disk in the vertical direction, then we could treat the oscillation in the vacuum limit. The probability of the flavor transition from neutrino $i$ ($\nu_{i}$) to neutrino ${j}$ ($\nu_{j}$) 
is given by
\begin{equation}
P_{\nu_{i}\rightarrow \nu_{j}} = -4\sum_{\alpha>\beta}\Re(U^*_{i \alpha}U_{j \alpha}U_{i \beta}U^*_{j \beta})\sin^2(1.27\Delta m^{2}_{\alpha \beta}\frac{d_{k}(\textrm{km})}{\epsilon^{k}_{\nu_{i}}(\textrm{GeV})}) 
+2\sum_{\alpha>\beta}\Im(U^*_{i \alpha}U_{j \alpha}U_{i \beta}U^*_{j \beta})\sin(2.54\Delta m^{2}_{\alpha \beta}\frac{d_{k}(\textrm{km})}{\epsilon^{k}_{\nu_{i}}(\textrm{GeV})}),
\label{eq:14}
\end{equation}
where $\Delta m^{2}_{\alpha \beta} = m^{2}_{\alpha} - m^{2}_{\beta}$ is the neutrino mass-squared difference. 
The values adopted for the normal mass hierarchy, obtained from the global fit of neutrino oscillation parameters provided by the Particle Data Group (PDG) \citep{PDG}, are $\Delta m^{2}_{21} = 7.53 \times 10^{-5}$, $\Delta m^{2}_{32} = 2.453 \times 10^{-3}$, and $\Delta m^{2}_{31} = \Delta m^{2}_{21}+\Delta m^{2}_{32}$.

In the Standard Model, the violation of charge conjugation parity symmetry (CP-violation) is considered to be a significant factor in distinguishing the flavor transition of neutrinos from anti-neutrinos \citep{CPviolation,CP2}.
In this work, we compare two cases where the CP-violating phase ($\delta_{\text{CP}}$) in the PMNS matrix is varied to be $0^{\circ}$ and $245^{\circ}$.
The plausible value of $\delta \rm CP = 245^{\circ}$ is based on the recent global fit of the average CP-violating phase measured in atmospheric and accelarator experiments: T2K, NOVA, and SKAM \citep{PDG}. $\delta \rm CP = 245^{\circ}$ introduces an extra phase term resulting in the different transition probability between $\nu_{i} \rightarrow \nu_{j}$ and $\bar{\nu}_{i} \rightarrow \bar{\nu}_{j}$ \citep{CPviolation} as given by
\begin{equation}
     P_{\bar{\nu}_{i}\rightarrow \bar{\nu}_{j}}
    =
    P_{\nu_{i}\rightarrow \nu_{j}} -4\sum_{\alpha>\beta}\Im(U^*_{i \alpha}U_{j \alpha}U_{i \beta}U^*_{j \beta})\sin(2.54\Delta m^{2}_{\alpha \beta}\frac{d_{k}(\textrm{km})}{\epsilon^{k}_{\nu_{i}}(\textrm{GeV})}).
\label{eq:15}
\end{equation}

The transition probabilities (Equations \ref{eq:14}-\ref{eq:15}) are utilized to account for the change in the total number of neutrinos and anti-neutrinos of the initial flavor before the neutrino annihilation luminosity (Equation \ref{eq:11}) is calculated. We also ignore the effect of general relativity on survival probability. However, this may not affect the total annihilation luminosity when the survival probability is sensitive to $d_{k}\sim$ $10^2$--$10^3$ for neutrino with energy $\sim \mathcal{O}(1-10)$ MeV \citep{SwamiOsc}. Therefore, we assume that the luminosities of flavor $i$ of neutrino and anti-neutrino at the annihilation point from those cells are given by $ P_{\nu_{i}\rightarrow \nu_{i}}\cdot l^{k}_{\nu_{i}}$ and $P_{\bar{\nu}_{i}\rightarrow \bar{\nu}_{i}}\cdot l^{k'}_{\bar{\nu}_{i}}$. The rest of the particles turning into other flavors could transfer their initial energies into other annihilation processes corresponding to their final flavor states with luminosities $P_{\nu_{i}\rightarrow \nu_{j}}\cdot l^{k}_{\nu_{i}}$ and $P_{\bar{\nu}_{i}\rightarrow \bar{\nu}_{j}}\cdot l^{k'}_{\bar{\nu}_{i}}$. 

\section{RESULTS} \label{sec:result}

\begin{figure*}[ht]
\centerline{
\includegraphics*[width=0.5\textwidth]{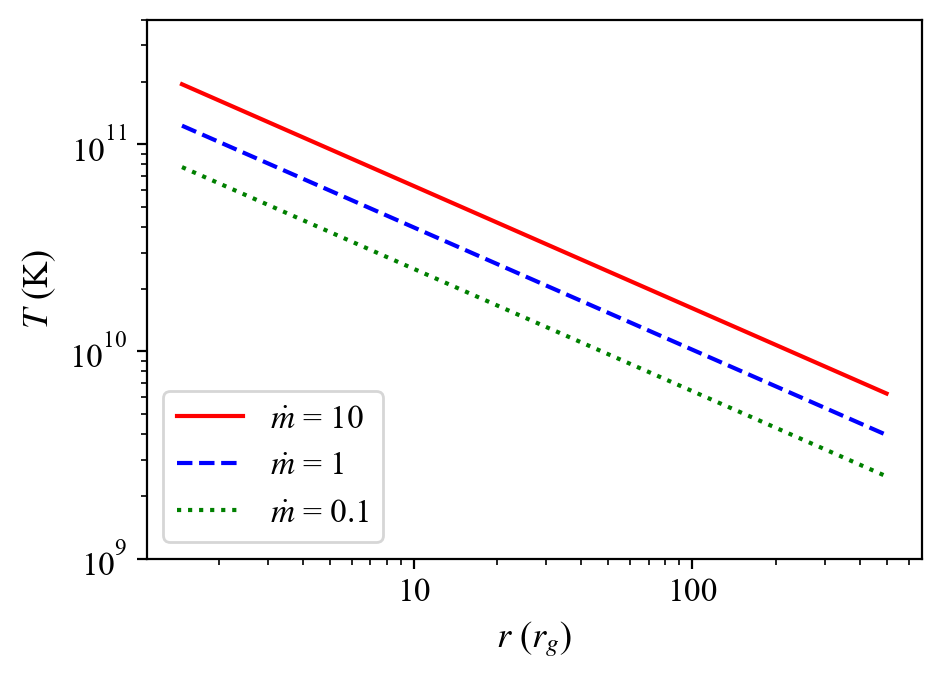}
\put(-60,150){$a$ = 0.99}
\includegraphics*[width=0.5\textwidth]{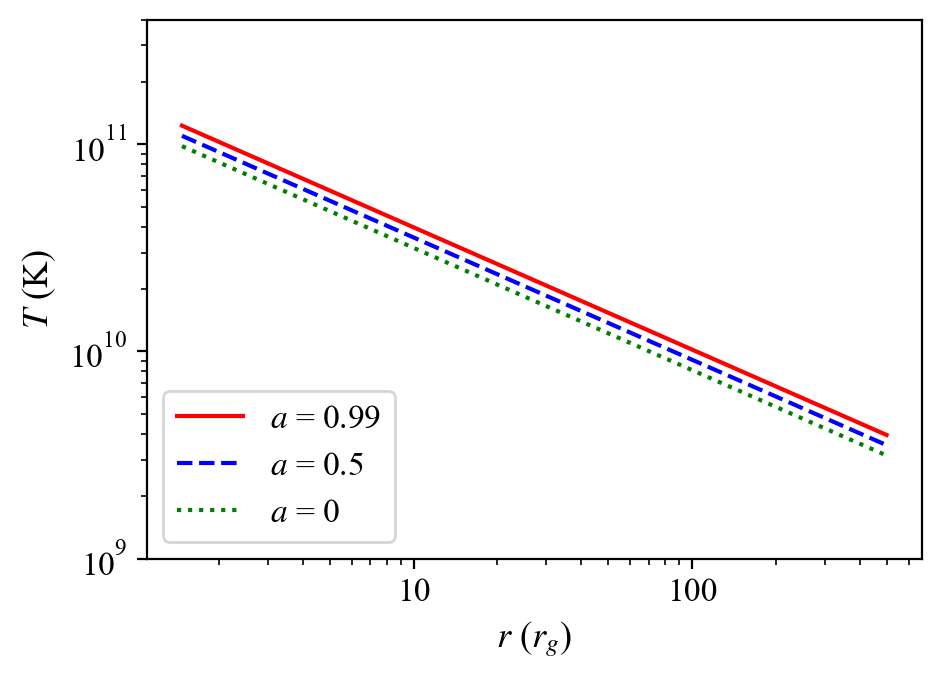}
\put(-60,150){$\dot{m}$ = 1}
\vspace{-0.15cm}
}
\caption{Disk temperature profiles for $a =0.99$ (left) when $\dot{m} =$ 0.1, 1, 10 (dotted, dashed, and solid lines respectively), and for $\dot{m} = 1$ (right) when $a =$ 0, 0.5, 0.99 (dotted, dashed, and solid lines, respectively).}
\label{fig:2}
\end{figure*}

Firstly, we calculate various thermodynamic quantities, including temperature (Equation \ref{eq:1}), matter density (Equation \ref{eq:3}), neutrino and anti-neutrino cooling rates (Equation \ref{eq:6}), and luminosity (Equation \ref{eq:10}) for different mass accretion rates $\dot{m} = 0.1, 1, 10$ and black hole spins $a = 0, 0.5, 0.99$. Figure \ref{fig:2} shows the temperature profiles of the accretion disk when the black hole mass and the half-thickness of the disk are fixed, as an example, at $M = 3$ $M_\odot$ and $H/r = 0.1$. As previously mentioned, our temperature profile is employed from the empirical solution derived by \cite{LiuDetect2016}, allowing us to infer the effects of $\dot{m}$ and $a$ on $T$ from the coefficients of Equation \ref{eq:1}.

As expected, higher temperature is observed in the cases of higher $\dot{m}$ and $a$, and the change in the temperature is more sensitive to variations in $\dot{m}$ than in $a$. 
Additionally, the disk temperature increases towards the innermost region closer to the central black hole.

\begin{figure*}[ht]
\centerline{
\includegraphics[width=0.5\textwidth]{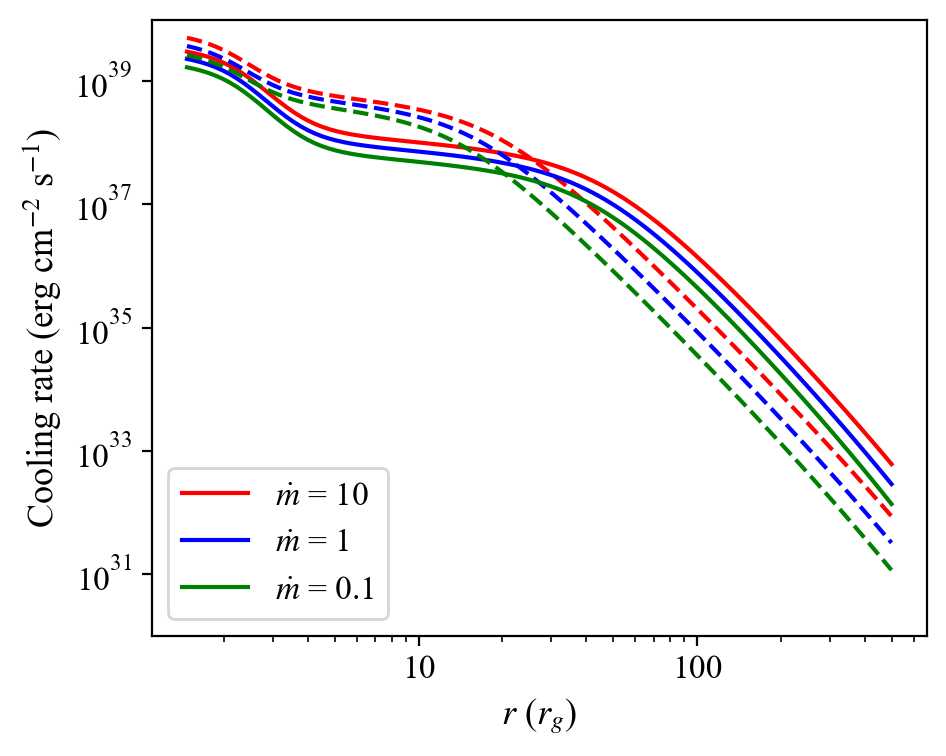}
\put(-60,170){$a$ = 0.99}
\includegraphics[width=0.5\textwidth]{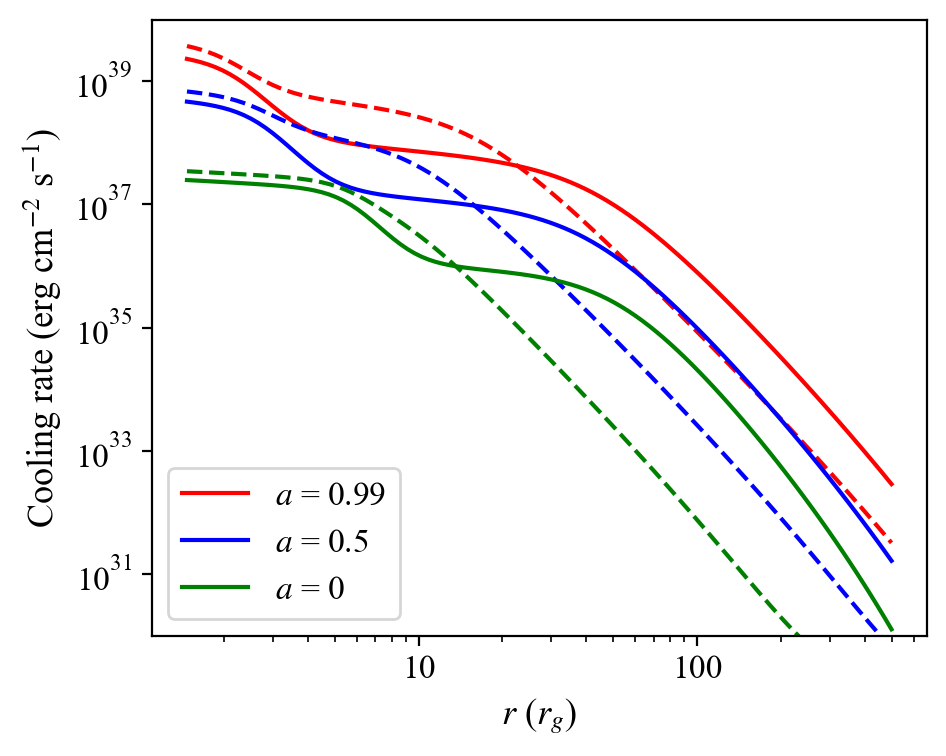}
\put(-60,170){$\dot{m}$ = 1}
\vspace{-0.2cm}
}
\caption{Radial profiles of neutrino cooling rate $Q_{\nu} = \sum_{i} Q_{\nu_{i}}$ (solid line) and anti-neutrino cooling rate $Q_{\bar{\nu}} = \sum_{i} Q_{\bar{\nu}_{i}}$ (dashed line) as a function of $r$ for $a = 0.99$ (left), when $\dot{m} =$ 0.1, 1, 10 (green, blue, and red color respectively), and for $\dot{m} = 1$ (right), when $a =$ 0, 0.5, 0.99 (green, blue, and red color respectively).}
\label{fig:3}
\end{figure*}

Figure~\ref{fig:3} illustrates how the neutrino cooling rate varies across the disk. The cooling rate scales with the disk temperature and significantly depends on the optical depth $\tau$, which links to the number density of electrons, protons, and neutrons. It is evident that the cooling rate also increases with $\dot{m}$ and $a$. Furthermore, in the inner region $r \lesssim 30~r_{\rm g}$ with $\tau \gg 1$, the cooling rate for anti-neutrinos exceeds that of neutrinos. This is due to the dominance of free nucleons in the dissociated matter, with the number densities of heavier nuclei being neglected. Therefore, the neutrino emission from cooling mechanism only depends on free nucleons. According to Equation \ref{eq:5}, the fixed $\eta_e$ results in the domination of neutrons in the region near a black hole, and hence the cooling process produces more anti-neutrino than neutrino in this region. For $r\gtrsim 30~r_{\rm g}$ with $0 <\tau < 1$, neutrinos dominate the disk, resulting in a higher cooling rate. The disk neutrino-dominated region becomes smaller for larger $a$ because of the rapid change in optical depth due to the thermal distribution. When $a$ change from 0 to 0.99, the disk neutrino-dominated region changes from the radial distance of 10--500 $r_{\rm g}$ to 30--500 $r_{\rm g}$.

Table \ref{tab:neutrino luminosity} shows the corresponding neutrino and anti-neutrino luminosities obtained by integrating the cooling rate over the disk emission areas when $a$ and $\dot{m}$ are varied. The total neutrino and anti-neutrino luminosity increase with $a$ and $\dot{m}$. When $0.1 \le \dot{m} \le 10$ and $0 \le a \le 0.99$, the total neutrino and anti-neutrino luminosity vary between 10$^{51}$--10$^{53}$ erg s$^{-1}$. However, we can see that $L_{\bar{\nu}}$ is slightly more than $L_{\nu}$ for $a =$ 0.99. This is because the anti-neutrino cooling rate has more efficiency near the black hole, $r \lesssim 30~r_{\rm g}$ (see Figure \ref{fig:3}, left panel). In cases of $a =$ 0, 0.5, the cooling rate of anti-neutrino is slightly greater than that of neutrino at the small radius of $r \lesssim 4~r_{\rm g}$, but the cooling rate of neutrino is significantly larger than that of anti-neutrino (see Figure \ref{fig:3}, right panel). This generally results in $L_{\nu} > L_{\bar{\nu}}$ for these cases.  

\begin{deluxetable}{cccc}[ht]
\tabletypesize{\small}
\tablewidth{\textwidth} 
\tablecaption{Neutrino(anti-neutrino) luminosity dependent on the spin parameter $a$ and the unitless mass accretion rate $\dot m$.}
\label{tab:neutrino luminosity}
\tablehead{
\colhead{$a$} & \colhead{$\dot{m}$} & \colhead{$L_{\nu}$}          & \colhead{$L_{\bar{\nu}}$}\\
\colhead{}    &                     & \colhead{($\rm erg~s^{-1}$)} & \colhead{($\rm erg~s^{-1}$)}} 
\startdata 
0.00  & 0.1 & 8.55$\times10^{51}$ & 4.16$\times10^{51}$ \\ 
0.00  & 1   & 1.41$\times10^{52}$ & 7.74$\times10^{51}$ \\ 
0.00  & 10  & 2.28$\times10^{52}$ & 1.42$\times10^{52}$ \\ 
0.50  & 0.1 & 5.88$\times10^{52}$ & 4.76$\times10^{52}$ \\ 
0.50  & 1   & 9.29$\times10^{52}$ & 8.31$\times10^{52}$ \\ 
0.50  & 10  & 1.45$\times10^{53}$ & 1.39$\times10^{53}$ \\ 
0.99  & 0.1 & 3.56$\times10^{53}$ & 3.72$\times10^{53}$ \\ 
0.99  &  1  & 5.52$\times10^{53}$ & 5.86$\times10^{53}$ \\ 
0.99  & 10  & 8.45$\times10^{53}$ & 9.01$\times10^{53}$ 
\enddata
\end{deluxetable}

The corresponding annihilation luminosity for all neutrino (anti-neutrino) flavors is shown in Table \ref{tab:Annihilation luminosity}. We compare the results with and without the effect of the neutrino oscillation. Here, the CP-violating phase is set to be $\delta \textrm{CP} = 0^{\circ}$. For all cases, the flavor transition in the vacuum results in a reduction of electron neutrino annihilation luminosity around 10--22\%. On the contrary, the muon neutrino luminosity increases by 6--45\%, 
while tau neutrino luminosity increases by 37--60\%. 
The reason is that the magnitude of the annihilation luminosity drops with $d^4$ (See Equation \ref{eq:11}) when the mean energy of neutrino (anti-neutrino) is around MeV. According to Equation \ref{eq:14}, a number of energetic $\nu_{e}$ are transformed with higher probability into $\nu_{\tau}$ than $\nu_{\mu}$. Note that the radiation is dominated by $\nu_{e}$, so that the total annihilation luminosity should be mainly affected by $\nu_{e}$.

\begin{deluxetable}{ccccccccc}[ht]
\tabletypesize{\small}
\tablecaption{Annihilation luminosity of neutrinos dependent on the spin parameter $a$ and the unitless mass accretion rate $\dot m$. Results are presented separately for each flavor denoted with the subscripts $e$, $\mu$ and $\tau$ for the parameters of electron-, muon-, and tau neutrino, respectively. We compare the cases of the total annihilation luminosity obtained when neutrino oscillation is neglected ($ L_{\nu\bar{\nu}}$) and when it is taken into account ($L^{\textrm{osc,0}}_{\nu\bar{\nu}}$). Their relative difference defined as $\Delta L^{\textrm{osc}}_{\nu\bar{\nu}}/ L_{\nu\bar{\nu}} = (L^{\textrm{osc,0}}_{\nu\bar{\nu}} - L_{\nu\bar{\nu}})/ L_{\nu\bar{\nu}}$ is also presented in the last column.}
\label{tab:Annihilation luminosity}
\tablehead{
\colhead{$a$} & \colhead{$\dot{m}$}& \colhead{$L_{\nu_{e}\bar{\nu}_{e}}$} & \colhead{$L_{\nu_{\mu}\bar{\nu}_{\mu}}$} & \colhead{$L_{\nu_{\tau}\bar{\nu}_{\tau}}$} & \colhead{$L^{\rm{osc,0}}_{\nu_{e}\bar{\nu}_{e}}$} & \colhead{$L^{\rm{osc,0}}_{\nu_{\mu}\bar{\nu}_{\mu}}$} & \colhead{$L^{\rm{osc,0}}_{\nu_{\tau}\bar{\nu}_{\tau}}$} & \colhead{$\Delta L^{\rm osc}_{\nu\bar{\nu}}/L_{\nu\bar{\nu}}$}\\
\colhead{} &  & \colhead{($\rm
erg~s^{-1}$)}& \colhead{($\rm erg~s^{-1}$)} & \colhead{($\rm erg~s^{-1}$)} & \colhead{($\rm erg~s^{-1}$)} & \colhead{($\rm erg~s^{-1}$)} & \colhead{($\rm
erg~s^{-1}$)} } 
\startdata 
0.00  & 0.1 & 3.83$\times10^{46}$ & 7.32$\times10^{45}$ & 7.32$\times10^{45}$ & 3.41$\times10^{46}$ & 7.83$\times10^{45}$ & 1.00$\times10^{46}$ & $-0.019$\\ 
0.00  & 1   & 1.92$\times10^{47}$ & 2.55$\times10^{46}$ & 2.55$\times10^{46}$ & 1.58$\times10^{47}$ & 2.82$\times10^{46}$ & 3.91$\times10^{46}$ & $-0.073$\\ 
0.00  & 10  & 8.71$\times10^{47}$ & 8.68$\times10^{46}$ & 8.68$\times10^{46}$ & 7.04$\times10^{47}$ & 9.79$\times10^{46}$ & 1.39$\times10^{47}$ & $-0.099$\\ 
0.50  & 0.1 & 1.47$\times10^{49}$ & 7.41$\times10^{47}$ & 7.41$\times10^{47}$ & 1.14$\times10^{49}$ & 8.95$\times10^{47}$ & 1.18$\times10^{48}$ & $-0.167$\\ 
0.50  & 1   & 5.12$\times10^{49}$ & 1.74$\times10^{48}$ & 1.74$\times10^{48}$ & 3.97$\times10^{49}$ & 2.16$\times10^{48}$ & 2.80$\times10^{48}$ & $-0.183$\\ 
0.50  & 10  & 1.57$\times10^{50}$ & 3.70$\times10^{48}$ & 3.70$\times10^{48}$ & 1.21$\times10^{50}$ & 4.95$\times10^{48}$ & 5.84$\times10^{48}$ & $-0.198$\\ 
0.99  & 0.1 & 1.17$\times10^{51}$ & 1.63$\times10^{49}$ & 1.63$\times10^{49}$ & 9.15$\times10^{50}$ & 2.32$\times10^{49}$ & 2.62$\times10^{49}$ & $-0.198$\\ 
0.99  &  1  & 2.73$\times10^{51}$ & 3.52$\times10^{49}$ & 3.52$\times10^{49}$ & 2.14$\times10^{51}$ & 5.12$\times10^{49}$ & 5.59$\times10^{49}$ & $-0.198$\\ 
0.99  & 10  & 5.76$\times10^{51}$ & 7.75$\times10^{49}$ & 7.75$\times10^{49}$ & 4.52$\times10^{51}$ & 1.10$\times10^{50}$ & 1.21$\times10^{50}$ & $-0.197$ 
\enddata
\end{deluxetable}

\begin{figure*}[ht]
\centerline{
\hspace{0.5cm}
\includegraphics[width=0.5\textwidth]{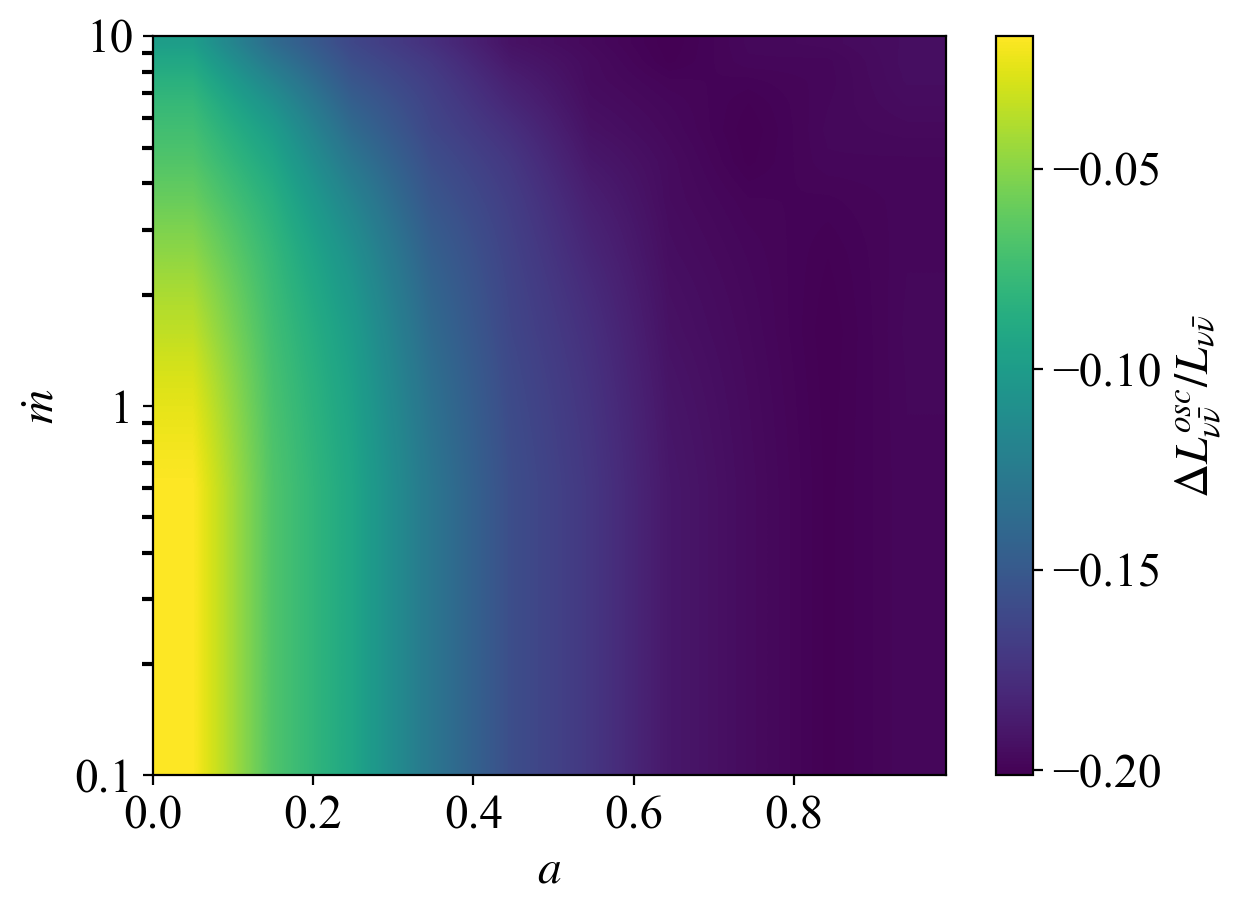}
\hspace{-0.7cm}
}
\caption{Relative annihilation luminosity difference $\Delta L^{\textrm{osc}}_{\nu\bar{\nu}}/ L_{\nu\bar{\nu}}$ varying with $a$ and $\dot m$. Note that we define $\Delta L^{\textrm{osc}}_{\nu\bar{\nu}}/ L_{\nu\bar{\nu}} = (L^{\textrm{osc,0}}_{\nu\bar{\nu}} - L_{\nu\bar{\nu}})/ L_{\nu\bar{\nu}}$ as the relative difference of the total annihilation luminosity between the cases when the effect of neutrino oscillation is included ($L^{\textrm{osc,0}}_{\nu\bar{\nu}}$) and excluded ($L_{\nu\bar{\nu}}$). The CP-violating phase is $\delta \textrm{CP} = 0$.}
\label{fig:4}
\end{figure*}

A global picture of the relative difference of the total annihilation luminosity with and without neutrino oscillation when $\delta \textrm{CP} = 0^{\circ}$ is illustrated in Figure \ref{fig:4}. Neutrino oscillation tends to decrease the annihilation luminosity by a factor of $\sim$ 0.01--0.20. The difference is more pronounced for higher $a$ and $\dot m$. Since the oscillation term depends on the ratio of the propagating distance per energy, $d/\epsilon$, an increase of $a$ and $\dot{m}$ would enhance the mean energies from local areas of the disk and broaden the oscillation pattern at the same propagating distance, hence reduce the flavor transition. However, $a$ and $\dot{m}$ affect the luminosity more than the flavor transition, resulting in a more negative $\Delta L^{\textrm{osc}}_{\nu\bar{\nu}}/ L_{\nu\bar{\nu}}$.

\begin{figure*}[ht]
\centerline{
\hspace{0.5cm}
\includegraphics[width=0.5\textwidth]{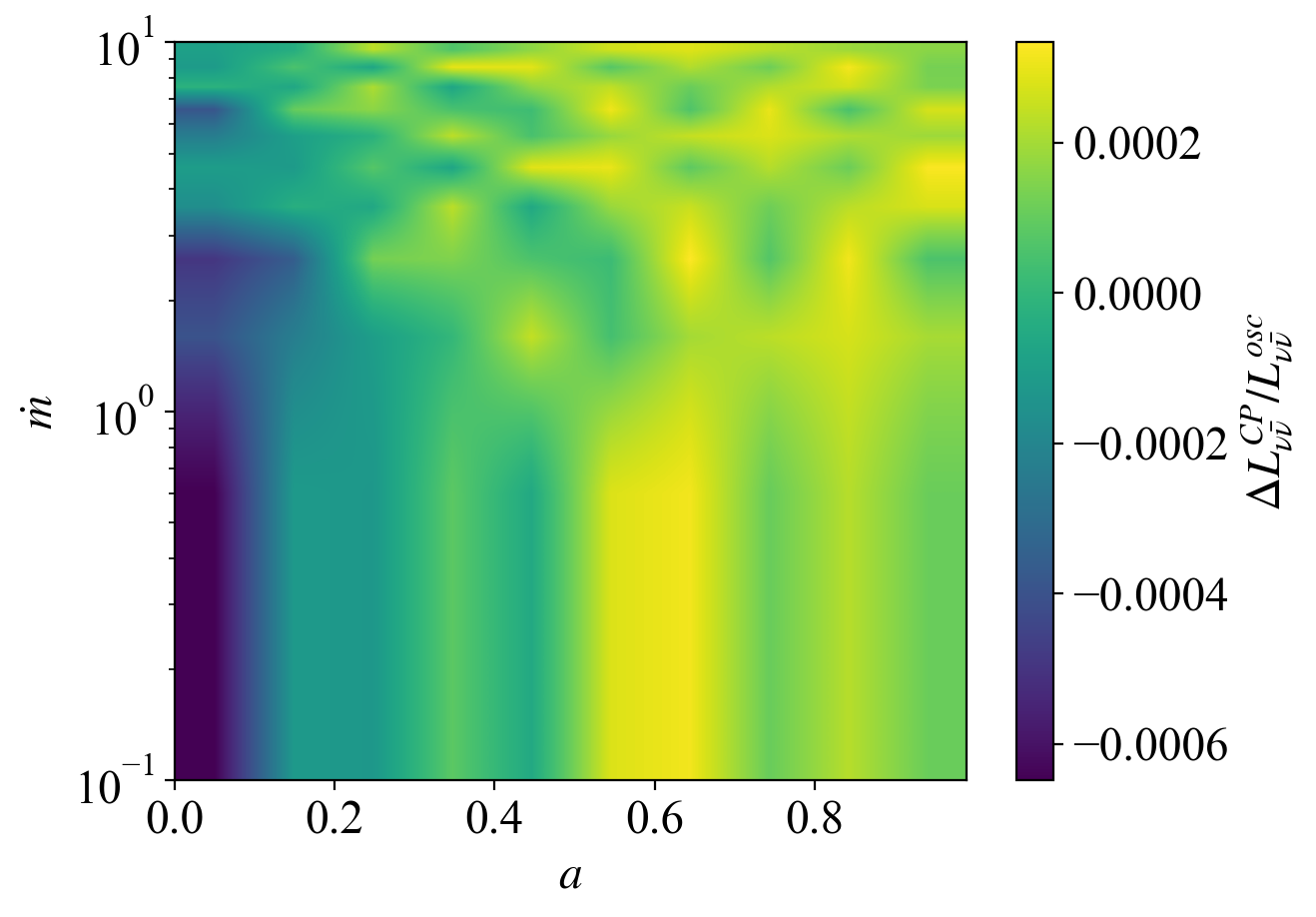}
\hspace{-0.7cm}
}
\caption{Relative annihilation luminosity difference $\Delta L^{\textrm{CP}}_{\nu\bar{\nu}}/ L^{\textrm{osc,0}}_{\nu\bar{\nu}} = (L^{\textrm{osc,245}}_{\nu\bar{\nu}} - L^{\textrm{osc,0}}_{\nu\bar{\nu}})/ L^{\textrm{osc,0}}_{\nu\bar{\nu}}$ varying with $a$ and $\dot m$. This figure highlights the effects of the CP-violating phase on the total annihilation luminosity where $L^{\textrm{osc,245}}_{\nu\bar{\nu}}$ and $L^{\textrm{osc,0}}_{\nu\bar{\nu}}$ refer to the cases when $\delta \textrm{CP} = 245^{\circ}$ and $\delta \textrm{CP} = 0^{\circ}$, respectively. Plus (red) and minus (blue) values indicate the case when $L^{\textrm{osc,245}}_{\nu\bar{\nu}} > L^{\textrm{osc,0}}_{\nu\bar{\nu}}$ and $L^{\textrm{osc,245}}_{\nu\bar{\nu}} < L^{\textrm{osc,0}}_{\nu\bar{\nu}}$, respectively. }
\label{fig:5}
\end{figure*}

Finally, we investigate the effect of the CP-violating phase by introducing the recent plausible value of $\delta \textrm{CP} = 245^{\circ}$. We calculate the relative difference of the annihilation luminosity $\Delta L^{\textrm{CP}}_{\nu\bar{\nu}}/ L^{\textrm{osc,0}}_{\nu\bar{\nu}} = (L^{\textrm{osc,245}}_{\nu\bar{\nu}} - L^{\textrm{osc,0}}_{\nu\bar{\nu}})/ L^{\textrm{osc,0}}_{\nu\bar{\nu}}$, where $L^{\textrm{osc,245}}_{\nu\bar{\nu}}$ and $L^{\textrm{osc,0}}_{\nu\bar{\nu}}$ represent the annihilation luminosity for $\delta \textrm{CP} = 245^{\circ}$ and $\delta \textrm{CP} = 0^{\circ}$, respectively. The result is shown in Figure \ref{fig:5}. Note that, according to our definition of $\Delta L^{\textrm{CP}}_{\nu\bar{\nu}}/ L^{\textrm{osc,0}}_{\nu\bar{\nu}}$, the plus (minus) value means that the annihilation luminosity in the cases of $\delta \textrm{CP} = 245^{\circ}$ is more (less) than when $\delta \textrm{CP} = 0^{\circ}$. As $a$ and $\dot{m}$ vary, the oscillation pattern changes in two ways simultaneously. Firstly, the addition of CP-violating phase can shift the oscillation pattern when the imaginary part in Equation \ref{eq:15} is nonnegligible. Secondly, the energies raised by increasing $a$ and $\dot{m}$ can slow down the flavor transformation, thereby broadening the oscillation pattern.

From Figure \ref{fig:5}, it is noticeable that the relative luminosity difference $\Delta L^{\textrm{CP}}_{\nu\bar{\nu}}/ L^{\textrm{osc,0}}_{\nu\bar{\nu}}$ alternates between negative and positive values in an approximate range of ($-6$ to 2)$\times 10^{-4}$ as $a$ and $\dot m$ vary. The amplitude of $\Delta L^{\textrm{CP}}_{\nu\bar{\nu}}/ L^{\textrm{osc,0}}_{\nu\bar{\nu}}$ appears to be larger for $a \lesssim 0.5$. This is because an increase in $a$ gradually raises the neutrino energy, resulting in a lower oscillation frequency. Then the chance for the flavor transition decreases, leading to a smaller effect from the CP-violating phase. 
Additionally, substantial variations in $\dot{m}$ are necessary to observe the fluctuation pattern. It is related to the variation in the peak of local energy and luminosity as $\dot{m}$ changes shown in Figure \ref{fig:4}. Since Figure \ref{fig:5} is in log scale for $\dot{m}$, the pattern becomes observable when $\dot{m} \gtrsim 1$. 

\begin{figure*}[ht]
\centerline{
\includegraphics*[width=0.5\textwidth]{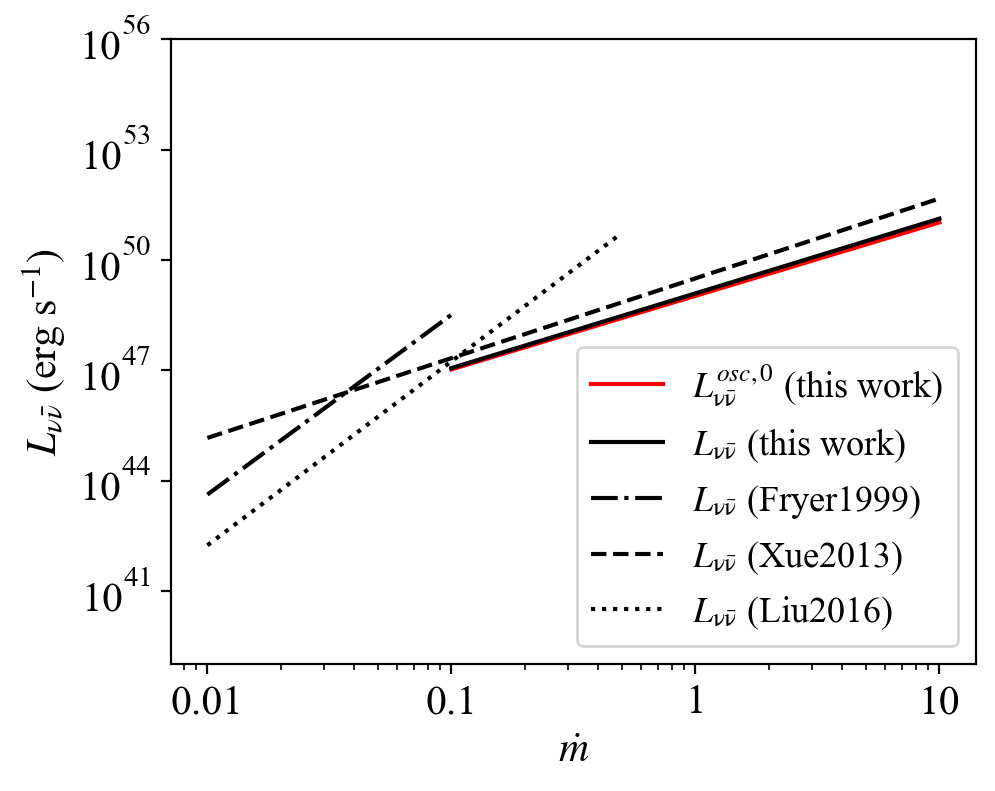}
\put(-190,170){$a$ = 0}
\includegraphics*[width=0.5\textwidth]{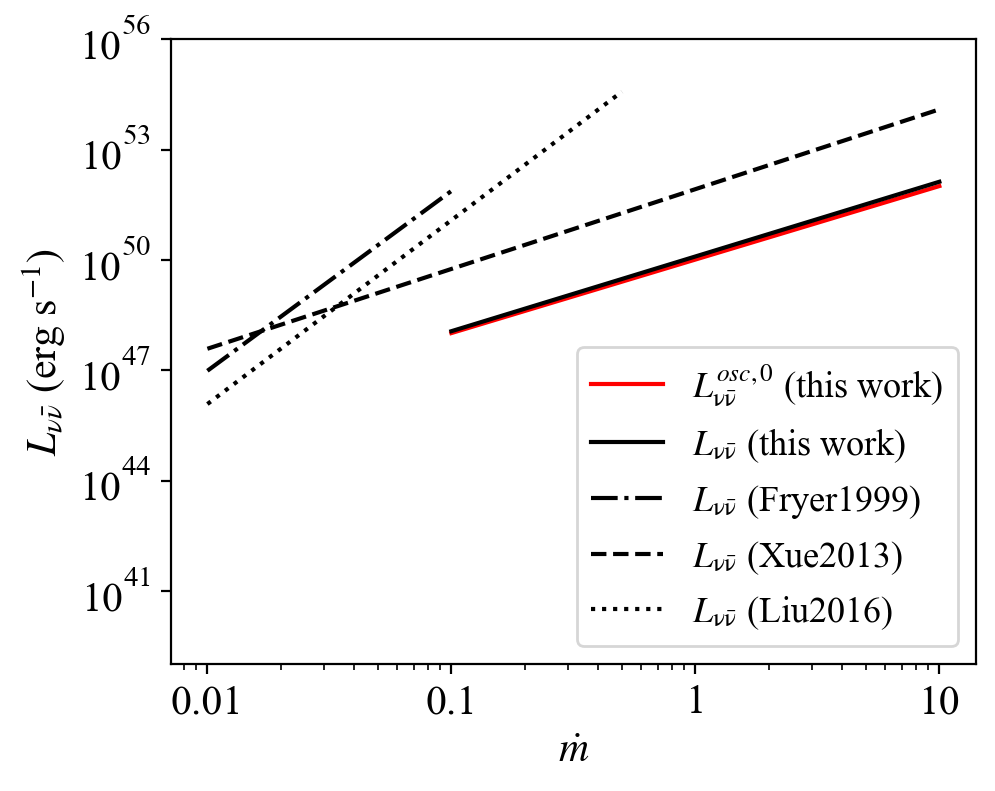}
\put(-190,170){$a$ = 0.99}
\vspace{-0.15cm}
}
\caption{Comparison of results in the annihilation luminosity  with neutrino oscillation $L^{\textrm{osc,0}}_{\nu \bar{\nu}}$ for $\delta \textrm{CP} = 0^{\circ}$ (red solid line), and without neutrino oscillation $L_{\nu \bar{\nu}}$ (black solid line), based on the Equations \ref{eq:16}--\ref{eq:17} derived from the fits with $a$ and $\dot{m}$ using linear regression method. The panels show the different fits between $a=0$ (left), and $a=0.99$ (right). Note that the annihilation luminosity for $\delta$CP = 0$^{\circ}$ is not much different from $\delta \textrm{CP} = 245^{\circ}$.
Results are compared
with the fits obtained by \cite{FryerFit} (dash-dotted line), \cite{XueNDAF} (dashed line), and \cite{LiuFit} (dotted line).}
\label{fig:6}
\end{figure*}

Last but not least, we derive equations of $ L_{\nu\bar{\nu}}$, and $L^{\textrm{osc,0}}_{\nu\bar{\nu}}$ as functions of $a$ and $\dot m$ from our model using linear regression fitting. The obtained relations are
\begin{equation}
    \log{L_{\nu\bar{\nu}}\,(\rm erg~s^{-1})} = 49.1 + 1.01a + 2.03\log{\dot{m}} \;,
    \label{eq:16}
\end{equation}
\begin{equation}
    \log{L^{\textrm{osc,0}}_{\nu\bar{\nu}}\,(\rm erg~s^{-1})} = 49.0 + 1.00a + 2.00\log{\dot{m}} \;.
    \label{eq:17}
\end{equation}
Note that the difference between the derived equations of $L^{\textrm{osc,245}}_{\nu\bar{\nu}}$ and $L^{\textrm{osc,0}}_{\nu\bar{\nu}}$ is very small (i.e., their coefficients are different only at the fifth decimal place). This allows for a direct comparison between our $L_{\nu\bar{\nu}}$ and those derived by other authors, such as \cite{FryerFit}, \cite{XueNDAF}, and \cite{LiuDetect2016}, as presented in the black lines in Figure~\ref{fig:6}. Notably, these authors did not consider the effect of neutrino oscillation in their calculations. The trend of increasing $L_{\nu\bar{\nu}}$ with $a$ and $\dot{m}$ is consistent, but the obtained amplitudes of $L_{\nu\bar{\nu}}$ are different. Additionally, in Figure~\ref{fig:6} we also show in red lines the derived relation for $L^{\textrm{osc,0}}_{\nu\bar{\nu}}$, corresponding to the case where neutrino oscillation with $\delta$CP = 0$^{\circ}$ is taken into account. Interestingly, the impact of neutrino oscillation on the annihilation luminosity appears to be less significant compared to the variations resulting from different assumptions made in the NDAF models used among different authors. 

\section{Discussion and Conclusion}\label{sec:con&discussion}

In this work, we investigate the neutrino/anti-neutrino luminosity and annihilation luminosity from the NDAF system by focusing on the effects of the neutrino oscillation above the accretion disk. The disk is assumed to be thin so that the thin disk approximation can be applied \citep[e.g.][]{PophamAccre}, and the analytic solution for the temperature profiles of the disk is utilized \citep{XueNDAF,LiuDetect2016}. The neutrino emission occurs at the ignition region of the disk where $1.473~{r_{\textrm{g}}} \leq r \leq 500~{r_{\textrm{g}}}$. We ignore nucleosynthesis when treating the disk consisting of dissociated matter. Under these conditions, the calculations for all disk profiles require significantly less time.

While electron degeneracy is a crucial factor for the thermal distribution \citep{KohriNDAF2002,KohriNDAF2005}, maintaining $Y_{e}$ in charge-neutrality constraints can be considered instead when $Y_{e}$ relates to the perfect equilibrium of neutrons and protons. Under this condition, thermal distribution will primarily rely on temperature, simplifying the luminosity calculation. Our luminosity profiles of the disk are well consistent with previous works, with the maximum luminosity approximately 10$^{53}$ erg s$^{-1}$ for an extreme accretion rate of $\dot{m} = 10$ \citep{PophamAccre,XueNDAF,LiuNDAF}.

Then, we investigate the annihilation luminosity with and without taking into account neutrino oscillation above the disk. When increasing either $a$ or $\dot{m}$, the neutrino emission (cooling rate) increases. This leads to an decrease in the total neutrino luminosity and annihilation luminosity. The annihilation luminosity without oscillation ranges between $\sim 10^{46}$--10$^{51}$ erg s$^{-1}$, depending on $a$ and $\dot{m}$. Note that we fix BH mass to be $M=3M_{\odot}$, but luminosity is expected to decrease with $M$ \citep[see, e.g., Equation 1 in][]{LiuFit}. Although, in this work, we fix $r_{\textrm{in}}= r_{\textrm{trp}}=$ 1.473 $ r_{\rm g}$ which is the intermediate value reported in \cite{XueNDAF}. We also explore the case when 
$r_{\textrm{in}}= 3$ $r_{\rm g}$ \citep{KohriNDAF2005,KawaNDAF}. The total annihilation luminosity is found to decrease by a factor of $\sim$ 2--3 compared to the result obtained when we fix $r_{\textrm{in}}=$ 1.473 $ r_{\rm g}$. Therefore, the choices of $r_{\textrm{in}}$ mentioned above do not significantly affect the order of magnitude of the obtained luminosity.

Referring to the pioneering work of \cite{PophamAccre}, \cite{FryerFit} utilized that result of annihilation luminosity and fitted it in the range of $\dot{m} = $ 0.01--0.1. Then, \cite{XueNDAF} performed a calculation with a more detailed understanding of neutrino physics. The results of this model are fitted to a wide range of accretion rates $\dot{m} =$  0.01--10, while the growth of the luminosity tends to decrease compared to \cite{FryerFit}. Later, \cite{LiuFit} extended the numerical solution of \cite{XueNDAF} by including the mass dependence and fitting it in the range of $\dot{m} = $0.01--0.5. The fitting line slope of annihilation luminosity is consistent with \cite{FryerFit}, but the luminosity increases more rapidly than \cite{XueNDAF}. The conclusion is that the increase in annihilation luminosity diminishes in high $\dot{m}$ due to neutrino trapping. In this work, we apply the calculation in \cite{XueNDAF} and fit the annihilation luminosity with $\dot{m} = $ 0.1--10 using linear regression. Our luminosity increases with $\dot{m}$ analogous to \cite{XueNDAF}, but we obtain a lower magnitude of $\sim$ 1 order. This is probably due to different assumptions used for electron degeneracy and neutrino trapping effects.

When taking into account the neutrino oscillation in a vacuum above the disk, the flavor transition reduces the magnitude of the annihilation luminosity by 1--19\% as $a$ and $\dot{m}$ increase. This is because the increase in energy leads to a lower oscillation frequency with less flavor transition, but is compensated by a rapid increase in luminosity. Specifically, transitions from muon flavor into other flavors occur with high probability. This results in the decrease of the electron neutrino annihilation luminosity by up to 22\%, and the increase of the muon- and tau-neutrino luminosities by up to 45\% and 60\%, respectively, as shown in Table~\ref{tab:Annihilation luminosity}. Additionally, in higher energy regimes (high $a$ and $\dot{m}$), the annihilation luminosity of muon flavor is comparable to tau flavor. This highlights the important role of energy, which depends on the specific setup of each NDAF model, in dictating neutrino oscillation behavior in this system.

At the same time, \cite{UribeOsc} found that the effect of oscillation inside the disk results in the flavor equipartition suppressing the electron neutrino cooling rate but increasing the muon- and tau-neutrino cooling rates. While the model presented in \cite{UribeOsc} predicts a higher annihilation luminosity compared to other previous works such as \citep{PophamAccre,XueNDAF,LiuFit}, they suggested a decrease in the total annihilation luminosity by as much as 80\% due to the neutrino oscillation inside the disk. Therefore, the net effect of neutrino oscillation from both inside and above the disk should induce a drastic decrease in the total annihilation luminosity. Based on the observed data, the majority of GRB luminosities exceed $\sim$ 10$^{51}$ $\textrm{erg s$^{-1}$}$ \citep{GRBLUM1}. Even if we consider the effect of neutrino oscillation, including both inside and above the disk, the energy transferred from annihilation into the relativistic electron-positron outflow may still be sufficient for GRB formation.

In a more realistic scenario, when thermal pressure surpasses gravity and inflow pressure, the mass could be lost from the disk into the outflow, similar to a large number of neutrinos and anti-neutrinos lost in neutrino emission. The densities of matters in the outflow should be considered for determining the interaction between the neutrinos (anti-neutrinos) and matters before and after annihilation \citep{LiCollecOsc,FastAccreOsc}. In this case, the probability of flavor oscillation should be separated into three channels: vacuum-dominated, resonance (high neutrino density), and matter-dominated (high matter density), depending on each region where neutrinos travel through. However, the matter-dominated has less impact on the oscillation above the disk since most energetic neutrinos (anti-neutrinos) annihilate in a very short distance \citep{OscGen}. Nevertheless, the neutrino annihilation should be taken into account the resonance effect because the oscillations are sensitive to the distances about 10 $r_{\rm{g}}$ $< d_{k} < $ 100 $r_{\rm{g}}$ for the system energies in the MeV range \citep{MatterS}. In our case, this could affect the number of annihilation events for each neutrino flavor near the rotating axis, but this may not change the trend of the result in which the annihilation luminosity decreases due to the reduction in the number of $\nu_{e}$, since this flavor dominates the total luminosity.

The T2K experiment identified CP-violation through the observation of muon flavor transforming into electron flavor for both neutrinos and anti-neutrinos \citep{CP2}. In our investigation, we explore scenarios where most muons transition into other flavors, including the CP-violating phase. However, our results suggest that changes in neutrino (anti-neutrino) energies play a more significant role than the CP-violating phase. This is attributed to the higher sensitivity of the oscillation frequency compared to the phase shift. Also, we found that the relative difference in annihilation luminosity between $\delta \textrm{CP} = 0^{\circ}$ and $\delta \textrm{CP} = 245^{\circ}$ is of the order of 10$^{-4}$; this suggests that the CP-violation in vacuum oscillation has minimal impact on neutrino annihilation. Consequently, detecting the effect of CP-violation on the formation of GRBs remains challenging.

Finally, we suggest that a more comprehensive model, exploring hydrodynamic flow and considering all possible interactions, should be considered. Improvements to the annihilation model could involve incorporating additional effects such as matter oscillation and neutrino-self interaction \citep{OscGen,MatterS,LiCollecOsc,FastAccreOsc} with realistic outflow. Additionally, the ray-tracing method \citep{LiuDetect2016} could be helpful in precisely tracking neutrinos and anti-neutrinos over distances. As mentioned in Section \ref{subsec:interpretation}, the interpretation of neutrino oscillation should be important for flavor transition in the NDAF model. This is because the interactions between neutrinos and antineutrinos with the disk can be considered in a variety of ways. Also, the CP-asymmetry of these particles leads to a different pattern of probability, and this effect accumulates in each step of the mechanism in the model. For future work, we aim to calculate all the possible flavor oscillations of the model, especially inside and above the disk.

\begin{acknowledgments}
\begin{center}
    ACKNOWLEDGEMENTS
\end{center}
This work was supported by the NSRF via the Program Management Unit for Human Resources \& Institutional Development, Research and Innovation (grant number B16F640076). CD acknowledges the support from the Institute for the Promotion of Teaching Science and Technology (IPST) of Thailand. PC thanks for the financial support from (i) Suranaree University of Technology (SUT), (ii) Thailand Science Research and Innovation (TSRI), and (iii) National Science, Research and Innovation Fund (NSRF, Fiscal Year 2024).

\end{acknowledgments} 

\appendix
\section{Neutrino emission}\label{sec:neutrinoemis}
The major cross sections from scattering off electrons, free protons, free neutrons and other elements particles are given by \citep{LiuNDAF},
\begin{equation*}
    \sigma_{e,\nu_{i}} = \frac{3k_{B}T\sigma_{0}E_{\nu_{i}}}{8m_{e}c^{2}}(1+\frac{\eta_{e}}{4})[(C_{V,\nu_{i}} + C_{A,\nu_{i}})^2
\end{equation*}
\begin{equation}
     + \frac{1}{3}(C_{V,\nu_{i}} - C_{A,\nu_{i}} )^2 ],
\end{equation}
\begin{equation}
    \sigma_{n_{1},\nu_{i}} = \frac{\sigma_{0}E^{2}_{\nu_{i}}}{4}[(C_{V,\nu_{i}} - 1)^2 + 3g^{2}_{A}(C_{A,\nu_{i}} - 1 )^2 ],
\end{equation}
\begin{equation}
    \sigma_{n_{2},\nu_{i}} = \frac{\sigma_{0}E^{2}_{\nu_{i}}}{4}\frac{1+3g^{2}_{A}}{4},
\end{equation}
\begin{equation}
    \sigma_{n_{j},\nu_{i}} = \frac{\sigma_{0}E^{2}_{\nu_{i}}}{16}(Z_{j} + N_{j})[1-\frac{2Z_{j}}{Z_{j} + N_{j}}(1-2\sin^{2}\theta_{W})]^2,
\end{equation}
where $\sigma_{0} = 4G^{2}_{F}(m_{e}c^{2})^2/\pi(\hbar c)^{4} \approx 1.71\times10^{-44}$ cm$^2$, $G_{F} \approx 1.436 \times 10^{-49}$ erg cm$^3$, $E_{\nu_{i}}$ is the mean energy of neutrinos in unit of $(m_{e}c^{2})$, $g_{A} \approx 1.26$, $\sin^{2}\theta_{W} \approx 0.23$, $Z_{j}$ and $N_{j}$ are defined as the number of proton and neutrons of a nucleus $X_{j}$, and $C_{V,\nu_{e}} = 1/2 + 2\sin^{2}\theta_{W}$, $C_{V,\nu_{\mu}} = C_{V,\nu_{\tau}} = -1/2 + 2\sin^{2}\theta_{W}$, $C_{A,\nu_{e}} = C_{A,{\bar{\nu}_{\mu}}} = C_{A,\bar{\nu}_{\tau}} = 1/2$, and $C_{A,\bar{\nu}_{e}} = C_{A,{\nu_{\mu}}} = C_{A,\nu_{\tau}} = -1/2$. 

\begin{equation}
    q_{\nu_{i}} = q_{\text{Urca}} + q_{\text{e}^{-} + \text{e}^{+} \rightarrow \nu_{i} + \bar{\nu}_{i}} + q_{n+n \rightarrow n+n+\nu_{i} + \bar{\nu}_{i}} + q_{\tilde{\gamma} \rightarrow \nu_{i} + \bar{\nu}_{i}}.
\end{equation}
First, $q_{\textrm{Urca}}$ is from the Urca process included in the proton-rich NSE \citep{XueNDAF}. This process plays an important role in neutrino radiation and relates only to electron neutrino. There are four major terms for electrons, positrons, free protons, free neutrons and nucleons \citep{LiuNDAF2007,KawaNDAF}, which are expressed by
\begin{equation*}
    q_{\textrm{Urca}} = q_{p+e^{-} \rightarrow n + \nu_{e}} + q_{n+e^{+} \rightarrow p + \bar{\nu}_{e}}
\end{equation*}
\begin{equation}
  + q_{n \rightarrow p + e^{-} + \bar{\nu}_{e}} + q_{X_{j}+e^{-} \rightarrow X'_{j} + \nu_{e}},
\end{equation}
with
\begin{equation*}
    q_{p+e^{-} \rightarrow n + \nu_{e}} = \frac{G^2_{F}\cos^{2}\theta_{c}}{2\pi^{2}\hbar^{3}c^{2}}(1+3g^{2}_{A})n_{1}
\end{equation*}
\begin{equation}
    \times \int^{\infty}_{Q}\sqrt{E^{2}_{e}-m^{2}_{e}c^{4}}(E_{e}-Q)^{3 }f(E_{e},\eta_{e})E_{e}dE_{e},
\end{equation}

\begin{equation*}
    q_{n+e^{+} \rightarrow p + \bar{\nu}_{e}} = \frac{G^2_{F}\cos^{2}\theta_{c}}{2\pi^{2}\hbar^{3}c^{2}}(1+3g^{2}_{A})n_{2}
\end{equation*}
\begin{equation}
    \times \int^{\infty}_{m_{e}c^{2}}\sqrt{E^{2}_{e}-m^{2}_{e}c^{4}}(E_{e}+Q)^{3 }f(E_{e},-\eta_{e})E_{e}dE_{e},
\end{equation}

\begin{equation*}
    q_{n \rightarrow p + e^{-} + \bar{\nu}_{e}} = \frac{G^2_{F}\cos^{2}\theta_{c}}{2\pi^{2}\hbar^{3}c^{2}}(1+3g^{2}_{A})n_{2}
\end{equation*}
\begin{equation}
    \times \int^{Q}_{m_{e}c^{2}}\sqrt{E^{2}_{e}-m^{2}_{e}c^{4}}(Q-E_{e})^{3 }[1-f(E_{e},\eta_{e})]E_{e}dE_{e},
\end{equation}

\begin{equation*}
    q_{X_{j}+e^{-} \rightarrow X'_{j} + \nu_{e}} = \frac{G^2_{F}\cos^{2}\theta_{c}}{2\pi^{2}\hbar^{3}c^{2}}g^{2}_{A}\frac{2}{7}N_{p}(Z_{j})N_{h}(N_{j})n_{j}
\end{equation*}
\begin{equation}
    \times \int^{\infty}_{Q'}\sqrt{E^{2}_{e}-m^{2}_{e}c^{4}}(E_{e}-Q')^{3 }f(E_{e},\eta_{e})E_{e}dE_{e},
\end{equation}
where $\cos^{2}\theta_{c} \approx 0.947$, $Q = (m_{n}-m_{p})c^{2}$, $Q' \approx \mu'_{n}-\mu'_{p}+\Delta$, $\mu'_{n}$ and $\mu'_{p}$ are the chemical potential of protons and neutrons in their own nuclei, $\Delta \approx 3\textrm{MeV}$ is the energy of the neutron $1f_{5/2}$ state above the ground state. $f(E,\eta)$ is the Fermi-Dirac distribution function given by
\begin{equation}
    f(E,\eta) = \frac{1}{e^{(E/k_{B}T)-\eta}+1},
\end{equation}
where $E$ is the energy in unit $(m_{e}c^2)$,
\begin{equation}
N_{p}(Z_{j}) = 
\begin{cases}
    0,               & Z_{j} < 20,\\
    Z_{j}-20,        & 20 < Z_{j} < 28,\\
    8,               & Z_{j} > 28,
\end{cases}
\end{equation}
\begin{equation}
N_{h}(N_{j}) = 
\begin{cases}
    6,               & N_{j} < 34,\\
    40-N_{j},        & 34 < N_{j} < 40,\\
    0,               & N_{j} > 40.
\end{cases}
\end{equation}
The second process is from the electron-positron pair annihilation into neutrinos $q_{\text{e}^{-} + \text{e}^{+} \rightarrow \nu_{i} + \bar{\nu}_{i}}$. This term can be neglected when electrons are in a degenerate state \citep{KohriNDAF2005}.

The third one is from the nucleon-nucleon bremsstrahlung $q_{n+n \rightarrow n+n+\nu_{i} + \bar{\nu}_{i}}$. The rate is the same for three flavors of neutrinos \citep{Itoh96,DiMatNDAF,LiuNDAF},
\begin{equation}
    q_{n+n \rightarrow n+n+\nu_{i} + \bar{\nu}_{i}} \approx 1.5\times 10^{27}\rho^{2}_{10}T^{5.5}_{11}\;\textrm{erg cm$^{-3}$s$^{-1}$}.
\end{equation}
The final main process is the plasmon decay $q_{\tilde{\gamma} \rightarrow \nu_{i} + \bar{\nu}_{i}}$, where plasmons $\tilde{\gamma} $ are photons interacting with electrons \citep{KawaNDAF},
\begin{equation*}
    q_{\tilde{\gamma} \rightarrow \nu_{e} + \bar{\nu}_{e}} = \frac{\pi^{4}}{6\alpha^{*}}C_{V,\nu_{e}}\frac{\sigma_{0}c}{(m_{e}c^{2})^{2}}\frac{(k_{B}T)^{9}}{(2\pi\hbar c)^6}\gamma^{6}
\end{equation*}
\begin{equation}
    \times(\gamma^{2} + 2\gamma + 2)\exp(-\gamma),
\end{equation}
\begin{equation*}
    q_{\tilde{\gamma} \rightarrow \nu_{\mu} + \bar{\nu}_{\mu}} = q_{\tilde{\gamma} \rightarrow \nu_{\tau} + \bar{\nu}_{\tau}} 
 = \frac{4\pi^{4}}{6\alpha^{*}}C_{V,\nu_{\mu}}\frac{\sigma_{0}c}{(m_{e}c^{2})^{2}}\frac{(k_{B}T)^{9}}{(2\pi\hbar c)^6}\gamma^{6}
\end{equation*}
\begin{equation}
    \times(\gamma^{2} + 2\gamma + 2)\exp(-\gamma),
\end{equation}
$\alpha^{*} \approx 1/137$ is the fine-structure constant and $\gamma \approx 5.565\times 10^{-2}[(\pi^{2}+3\eta^{2}_{e})/3]^{1/2}$.\\

\section{Oscillation parameters}\label{sec:oscparams}
Without considering the matter oscillation, self-interaction, and/or other interactions, the neutrino oscillation is calculated in the vacuum limit. If one assumes that the neutrino is not Majorana particle, the unitary transformation for any flavors is described by the Pontecorvo–Maki–Nakagawa–Sakata matrix (PMNS matrix) given by
\begin{equation}
U =
\begin{pmatrix}
c_{12}c_{23} & s_{12}c_{23} & s_{13}e^{-i\delta \textrm{CP}}\\
-s_{12}c_{23} -c_{12}s_{23}s_{13}e^{i\delta \textrm{CP}} & c_{12}c_{23}-s_{12}s_{13}s_{23}e^{i\delta \textrm{CP}} & c_{13}s_{23}\\
s_{12}s_{23} -c_{12}c_{23}s_{13}e^{i\delta \textrm{CP}}  & -c_{12}s_{23}-s_{12}s_{13}c_{23}e^{i\delta \textrm{CP}} & c_{13}c_{23}
\end{pmatrix},
\end{equation}
where $c_{\alpha\beta} = \cos{\theta_{\alpha\beta}}$, and $s_{\alpha\beta} = \sin{\theta_{\alpha\beta}}$, and the values $\theta_{12}$ = 33.65$^{\circ}$, $\theta_{23}$ = 47.64$^{\circ}$, and $\theta_{13}$ = 8.53$^{\circ}$, are taken from PDG.


\bibliography{sample631}{}
\bibliographystyle{aasjournal}



\end{document}